\documentclass[preprint,nofootinbib,aps,showpacs,showkeys]{revtex4}
\usepackage{amsmath}
\usepackage{amsfonts,amssymb}
\usepackage{graphicx}

\begin{document}
\title{Description of resonances in light nuclei using a microscopic cluster model}

\author{Theodoros Leontiou}
\email{Leontiou@Theory.phy.umist.ac.uk}
\author{Niels R. Walet}
\email{Niels.Walet@manchester.ac.uk} \affiliation{School of
Physics and Astronomy, The University of Manchester, PO Box 88,
Manchester, M60 1QD, UK}

\begin{abstract}
We investigate resonances in light halo nuclei using a fully
microscopic cluster model and the complex scaling method. We make
use of the hermitian representation of the complex scaling method.
The general structure of the cluster model is that of a
correlation operator acting on a starting function that describes a
number of neutrons relative to an alpha-particle. The correlation
operator is expanded in terms of a small non-orthogonal set of
Gaussian basis states and we make use of a simplified, central but
state-dependent, interaction. The many-body integrals required for
the computation are evaluated by a variational Monte-Carlo
algorithm. We show how to obtain resonant states for both $^5$He
and $^6$He.
\end{abstract}
\pacs{21.45.+v,21.60.Gx,21.60.Ka,24.30.Gd}
\keywords{resonance, complex scaling, microscopic, cluster, Monte Carlo}
\maketitle
\newpage


The study of resonant structures in loosely bound systems, such as
halo nuclei, is an interesting area of nuclear physics that promises
to be a rich source of information on many-body dynamics. There exists
a variety of methods for analyzing such structures and in this work we
consider an approach that belongs to the category of cluster models,
which have proven to be powerful tools in analyzing the structure of
such loosely bound systems. The most common way in which present
cluster models describe the continuum structure of light nuclei is by
treating only the most important degrees of freedom properly, and
ignore the difficulties with the Pauli principle that entails.  In
contrast to such approaches, we employ a fully microscopic
cluster model in order to provide a variational approximation to the
Schr\"odinger equation with, in this case, a slightly simplified
nucleon-nucleon interaction.

A related microscopic cluster model that can deal with three-body
resonances by complex-scaling in coordinate space was developed by
Csoto and applied to $^6$He, $^6$Li and $^6$Be
\cite{csoto:csoto1}. This model can account in many cases both for
the correct nuclear physics and proper three-body dynamics. Among
other results the authors predict the non-existence of the soft dipole
resonance in $^6$He. Another example of a similar study of halo nuclei
by the complex scaling method through a microscopic model can be found
in Ref.~\cite{kato}. An alternative to the use of the complex scaling
in coordinate space is to perform a contour deformation in momentum
space \cite{hagen}. This was applied at the level of a two-body
problem and an extension to few-body Borromean halo systems is in
progress.

Despite the success of the two- and three-body microscopic cluster
models the existing fully microscopic ones have problems when
considering resonances  \cite{csoto:csoto2}. The ab-initio
approaches, e.g., the Monte-Carlo approach of Ref.~\cite{wiringa},
even though very well suited for the bound-state problem, have
difficulties describing resonances.

By using the complex scaling method \cite{reinhardt,moiseyev:mo2} in
the `direct approach', our model can handle resonances correctly in a
straightforward way. The wavefunction we employ is fully microscopic
and is explicitly antisymmetrized so that the Pauli exclusion
principle is exactly satisfied. The general structure is that of a
correlation operator acting on a reference function that describes a
number of neutrons relative to an alpha-particle. We include a
multiplicative Jastrow factor for the short-range correlations and an
additive configuration-interaction (CI) term for the long-range
correlations. This last term only introduces pair correlations, whereas 
the Jastrow term also includes higher correlations.  This mechanism has
proved to be efficient in describing nuclear structure in various
applications to closed and open shell systems, where all particles are
treated on an equal footing \cite{buendia,niels}. Recently, this has
also been applied \cite{sarsa} to nuclei with $12\leq A\leq16$.

One advantage of the Jastrow-CI scheme is the explicit inclusion of
translational invariance and the fact that the correlation operators
can be sufficiently well approximated by a small non-orthogonal basis
set that makes application to heavier nuclei possible. For the moment
we confine ourselves to central V4 interactions and in particular we
make use of the S3 interaction of Afnan and Tang \cite{s3}. For the
computation of the many-body matrix elements we make use of the
Variational Monte Carlo method, which we have analyzed in detail
\cite{leontiou} as applied to this cluster model.

In this letter we focus on the cases of $^5$He and $^6$He. Although
experimentally $^5$He is unbound while $^6$He is lightly bound, the
absence of spin-orbit force in our simplified force results in both
nuclei being unbound. Nevertheless, both nuclei exhibit a rich
resonance structure. The current work thus provides an excellent
testing ground for the study of resonances, even though the nuclear
physics may not be wholly realistic.



There exists many different version of the complex scaling technique.
In the original complex-coordinate method (see, e.g.,
Ref.~\cite{reinhardt}) the resonance position $E_R$ and width $\Gamma$ 
are calculated from  the complex eigenvalue $E=E_R-i\Gamma/2$ of
a non-hermitian Hamiltonian. The non-hermitian Hamiltonian is
obtained from the original one by a transformation of the
particle coordinates. Here we shall only use the most straightforward
one, $r\rightarrow r\exp(i\theta)$, where the angle $\theta$
is the scaling parameter. Such a transformation leaves the bound
states unaffected and rotates the positive energy scattering states by
an angle of $2\theta$ into the lower part of the complex plane
relative to the threshold
\cite{reinhardt,moiseyev:mo2}. Resonant states $\Psi_{\rm res}$ appear as
complex eigenvalues in the lower half complex plane that are invariant
with respect to changes in the scaling parameter $\theta$, as long as this angle is large enough to uncover the resonances. [A more detailed review of
the complex scaling method can be found in \cite{moiseyev:mo2}.]

For a complex scaled Hamiltonian, the complex variational principle,
unlike the conventional one, is a stationary principle rather than a
minimum principle for either the resonance position or width. This is
not very attractive, and Moiseyev \cite{moiseyev:mo1} has formulated a
representation of the complex-coordinate method in which the resonance
position $E_r$ and width $-2E_i$ are variational parameters of a
hermitian Hamiltonian which gives additional stability, especially
when working with small basis sets, as in this work. It is also gives
a simple algorithm to  follow the path of a given eigenvalue as $\theta$ changes. 
The variational method was reformulated by Bylicki \cite{bylicki}, 
into a simpler and more convenient form.

In this method, we first rewrite the complex scaled Schr\"odinger
equation as
\begin{equation}\label{eq1}
(\hat{H}_r+i\hat{H}_i)(\Psi_r+i\Psi_i)=(E_r+iE_i)(\Psi_r+i\Psi_i),
\end{equation}
where $\hat{H}_r$ and $\hat{H}_i$ are the real and imaginary parts of
the scaled Hamiltonian $\hat{H}(re^{i\theta})$. In matrix form this becomes
\begin{equation}\label{eq2}
{\cal H}(E)\Psi=0,~~~
{\cal H}=\left(\begin{array}{cc}
-\hat{H}_i+E_i&\hat{H}_r-E_r\\
\hat{H}_r-E_r&\hat{H}_i-E_i
\end{array}\right),~~~\Psi=\left(\begin{array}{c}\Psi_i\\ \Psi_r\end{array}\right).
\end{equation}

Since the exact eigenvalue of $\hat{H}(re^{i\theta})$ is unknown we
can consider the alternative equation
\begin{equation}
{\cal H}(E)\Phi=\lambda\Phi,
\end{equation}
where $\Phi$ is an approximation to the exact wavefunction $\Psi$.
For a particular value of the scaling parameter $\theta$ the {\em
best} choice of $\Phi$ is obtained by minimizing the size of
$\lambda$, and $\Phi=\Psi$ if $\lambda=0$. Within this
approximation the parameters $E_r$ and $E_i$ that appear in ${\cal
H}$ no longer represent the exact real and imaginary parts of the
complex energy but become variational parameters. Thus in addition
to the linear variational principle that is employed for the
wavefunction we also have to vary $E_r$ and $E_i$ so as to
minimize $|\lambda|$.  Although $\lambda$ has both positive and
negative values one interesting property of Eq.~(\ref{eq2})
is that the spectrum of eigenvalues is symmetric around 0, thus if
$|\lambda|$ is an eigenvalue so is $-|\lambda|$.

Furthermore, the equation ${\cal H}^2\Phi=\lambda^2\Phi$ is useful
for obtaining bounds for the resonance position and width and it
can be used in an iterative method for determining the values of
$E_r$ and $E_i$ that minimize $|\lambda|$ \cite{moiseyev:mo1}.


The many body Hamiltonian is of the form
\begin{equation}
\hat{H}=\sum_{i=1}^A\frac{\hbar^2}{2m_i}\nabla^2_i+\sum_{1\leq i<j}^AV(ij),
\end{equation}
In this letter, we use a semi-realistic central nucleon-nucleon interaction 
(V4) interaction that has the general form
\begin{equation}
V(ij)=V_0(ij)+V_\sigma(ij)+V_\tau(ij)+V_{\sigma\tau}(ij).
\end{equation}
Here the $V_{\sigma/\tau}$  are the spin/isospin-dependent part composed of
spin/isospin-exchange operators. The radial part of each channel potential
is expanded as a sum of Gaussian functions. We use the S3 interaction of Afnan and Tang \cite{s3}. 

For the trial wavefunction we consider the linearized approximation of
the many-body wavefunction in terms of correlation operators acting
on an uncorrelated reference function \cite{bish}. This provides a
translationally invariant description of the many-body problem and was
applied to a number of closed-shell systems in terms of an
alpha-cluster model \cite{moliner}.  The wavefunction $\Psi$ is given
by $\Psi=\hat{F}\Phi_0,$ where $\Phi_0$ is the reference function,
that we take to be the product of the different cluster
wavefunctions. The correlation operator $\hat{F}$ is of the form
\begin{eqnarray}
\hat{F}=\sum_{k=1}^N \hat{F}_k\left(\sum_{k=1}^N a_k \sum_{i<j}f^k(ij)\right)\prod_{i<j}g(ij),\\
f(ij)=\exp(d_k r_{ij}^2),~~~g(ij)=1-k_1\exp(-\lambda_1r_{ij}^2)-k_2\exp(-\lambda_2r_{ij}^2). \label{eq:correlationop}
\end{eqnarray}
This type of wavefunction (referred to as the J-TICI(2) scheme) has been
extensively used for the alpha-particle
\cite{buendia,niels} 
where it was shown to provide an adequate description for the
ground-state properties. An important aspect of this approximation is
the expansion of the correlation operator in terms of the
non-orthogonal Gaussian functions $\exp(d_k r_{ij}^2)$. Only a very
small number of components is required for a reasonable convergence,
something that makes calculations in the J-TICI(2) scheme less
expensive than other microscopic methods.

Due to the complexity of the correlation operator we can make use of
the variational Monte Carlo method for the evaluation of the matrix
elements, where the probability density function $w$ is taken to be
the square of one of the components in the expansion of the
wavefunction, i.e., 
$
w=(\hat{F}_0\Phi_0)^2.
$
This is a natural choice for the PDF since the wavefunction can always
be written as the product of $w$ and a function $\Psi'$. This is
discussed in detail in Ref.~\cite{leontiou}.

Thus Eq.~(\ref{eq1}) is approximated in terms of a non-orthogonal
expansion. The non-orthogonality requires that in Eq. (\ref{eq2}) the
real and imaginary parts of the complex scaled Hamiltonian are
modified according to
\begin{equation}
\hat{H}_r\rightarrow N^{-1/2}\hat{H}_rN^{-1/2},~~~
\hat{H}_i\rightarrow N^{-1/2}\hat{H}_iN^{-1/2}.
\end{equation}
$N$ represents the overlap matrix and
$N^{-1/2}=C\Lambda^{-1/2}C^{-1}$, where $C$ is the matrix with the
eigenvectors of $N$ as its columns and $\Lambda$ is the diagonal
matrix with the eigenvalues of $N$ as its elements.


\begin{figure}
\centerline{\includegraphics[width=10cm,clip]{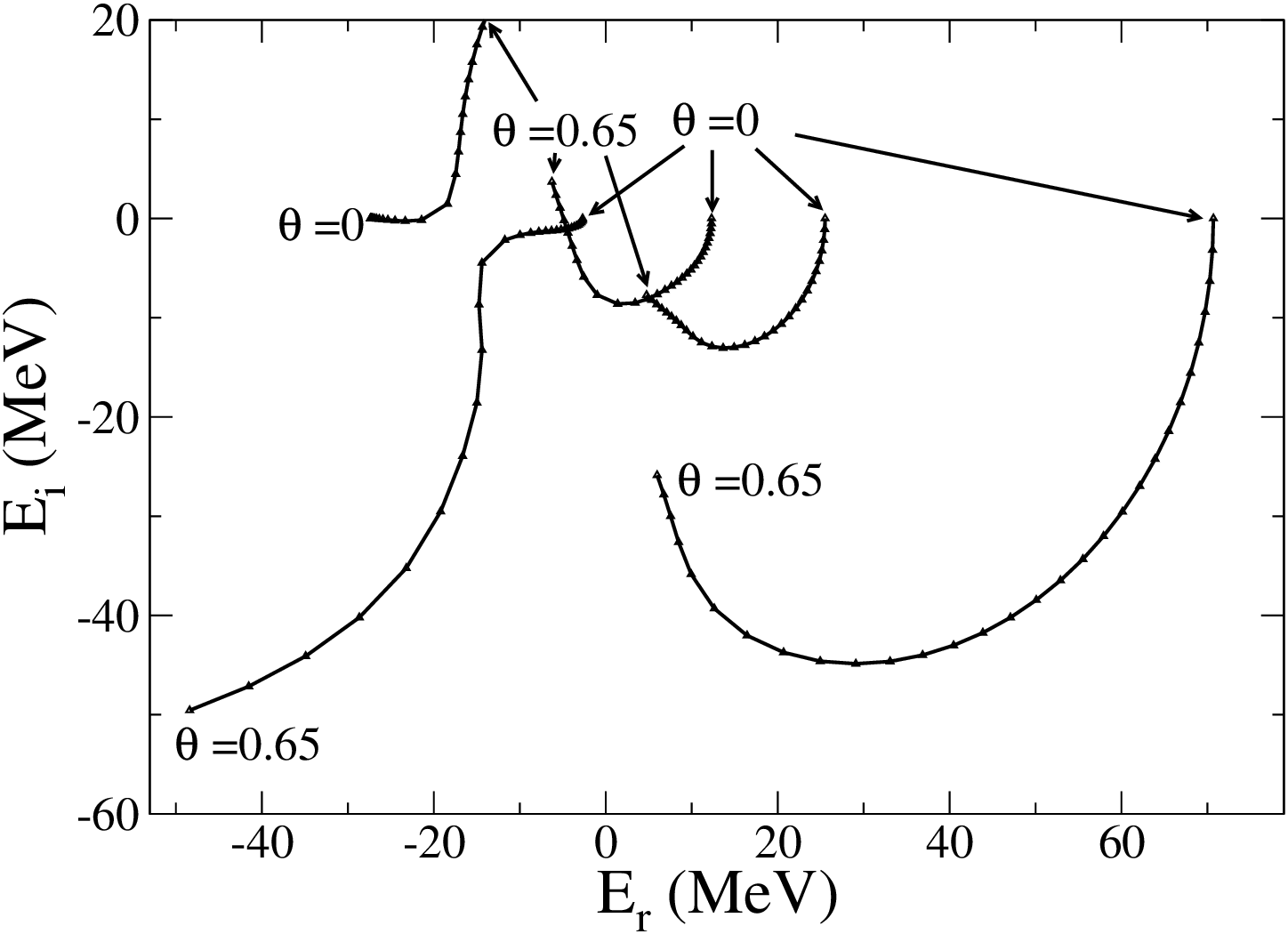}}
\caption{\label{fig1}  $\theta$-trajectories
for the spectrum of the alpha-particle for various values of the
rotation parameter $\theta$ between 0 and 0.65 radians. The value
of the ground state corresponding to $\theta=0$ is -27.3 MeV. The
threshold is not stationary but rotates into the upper quadrant of
the complex plain, while the rest of the eigenvalues move
clockwise into the lower complex plain.}
\end{figure}

We firstly examine the application of the formalism of Moiseyev and
Bylicki to the alpha-particle, where we do not expect to find any
resonances. For the alpha-particle reference function we use the
spherically symmetric harmonic oscillator ground state. It is an easy
numerical calculation both in terms of computation time and
complexity. We can go one step further by introducing one or two
additional nucleons, corresponding to the cases of $^5$He and $^6$He.

Experimentally the alpha-particle is a stable nucleus with 20 MeV
difference between the ground and first excited state. Searching for
resonances is thus not extremely important, but the calculation will
help to highlight the implications of the approximation scheme
used. The results are displayed in Fig.~\ref{fig1}.  The threshold for
this calculation is the ground state energy as obtained from the
variational approximation ($\approx$~-27.3 MeV). As can be seen from
Fig.~\ref{fig1} the ground-state remains almost constant for small
$\theta$ but when we increase $\theta$ further it eventually
moves. This is not unexpected since the variational ground state is
only an approximation to the exact threshold and it has a non-zero
overlap with continuum states. It also suggests that we should
optimize our basis to change with $\theta$---in this context we are
especially concerned about the Jastrow factor. The rest of the
eigenvalues move clockwise into the lower complex plane, eventually
turning up as well. The behaviour of the threshold is somewhat of a
worry, and shows that our variational form is not too good for finite
$\theta$. In all likelihood this is caused by the Gaussian functions
in the Jastrow correlations (i.e., the short range behaviour of the
wave function), but this requires confirmation. Clearly, we expect
all calculations to fail for $\theta=\pi/4$, where all Gaussians
become oscillatory.

Experimentally $^5$He is unbound by 0.798 MeV and is observed as a
$J^\pi=\frac{3}{2}^-;T=\frac{1}{2}$ resonance in the neutron
scattering on $^4$He. In our approximation the case of $^5$He is that
where a neutron is added to the alpha particle wavefunction, where the
additional coordinates are specified relative to the alpha-particle
center-of-mass. The additional neutron is placed within a spherical
shell and the many-body wavefunction has the form
\begin{eqnarray}
\Psi_{^5He}&=&\hat{F}{\cal A}\{
\Phi_\alpha f(r_{a5}){\cal Y}^L_M(r_{a5})\times\chi_{\sigma\tau}(S,T) \} ,\\
f(r_{a5})&=&\exp\left(-\left(\frac{r_{a5}-d}{w}\right)^2\right),
\end{eqnarray}
where ${\cal Y}^L_M(r_{a5})$ is a solid harmonic, while
$\chi_{\sigma\tau}(S,T)$ represents the spin and isospin degrees of
freedom. In the function $f(r_{a5})$ the parameter $d$ represents the
distance from the alpha-particle, while $w$ represents the width of
the shell. $r_{a5}$ is the relative distance of the neutron from the
alpha-particle. The linear correlation operator is invariant under
permutation of particle labels and can be taken outside the
antisymmetrizer.

\begin{figure}
\centerline{\includegraphics[width=8cm,clip]{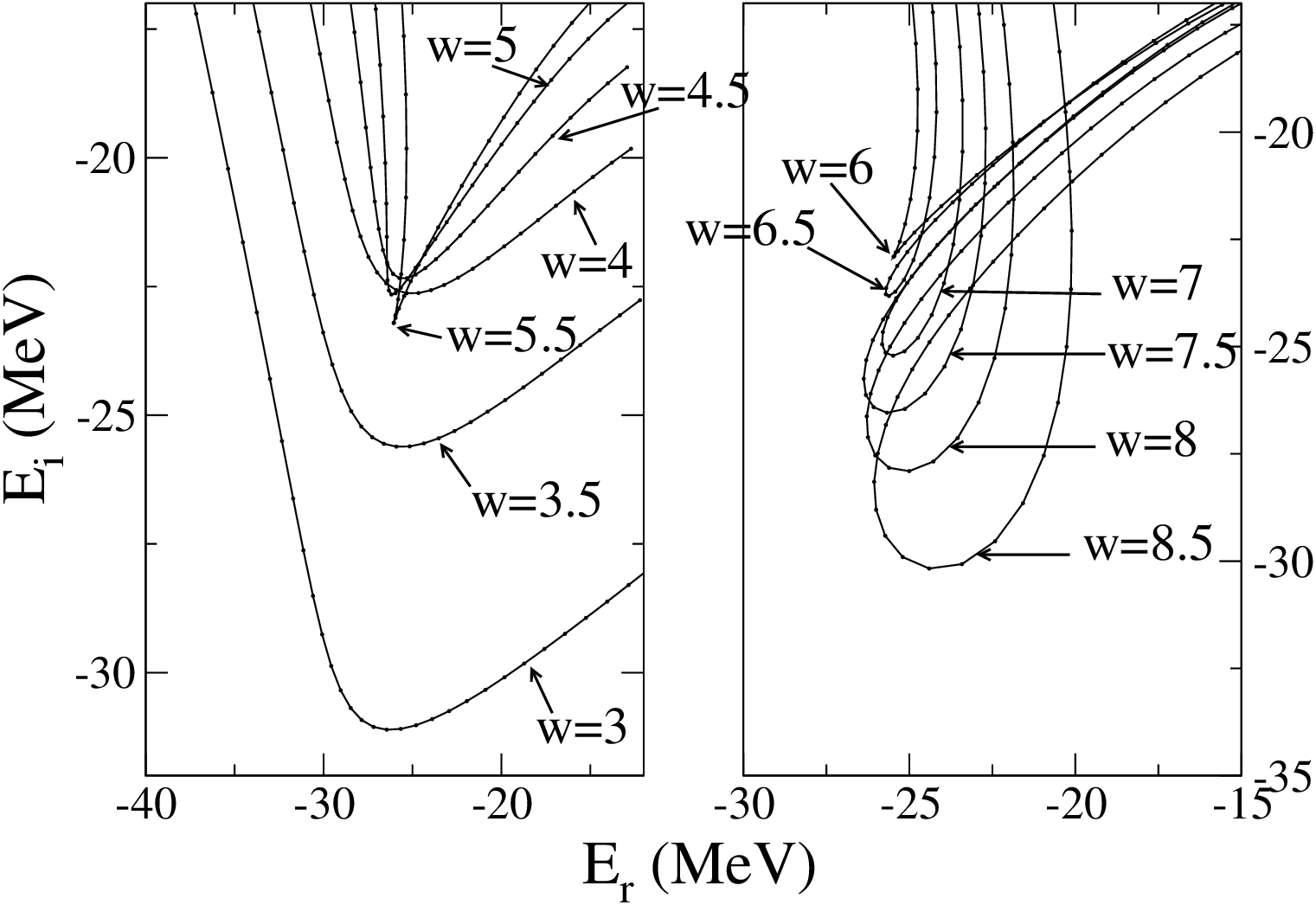}}
\caption{\label{fig2} $\theta$-trajectories obtained
for $^5$He $L=1$ ($J^\pi=1/2^-,3/2^-$) for various values of the width
parameter $w$. In each plot the rotation parameter $\theta$ is varied
in equal steps and the $\theta$-trajectories evolve from right to left
as $\theta$ is increased.}
\end{figure}

Both $d$ and $w$ are variational parameters and in principle we can
have a number of them, i.e. expand over a number of shells each placed
at a different distance from the alpha-particle with a given
width. However, as $w$ increases the value of $d$ becomes
insignificant and we can consider a single shell provided $w$ is large
enough. The results obtained for $L=1$ ($J^\pi=1/2^-,3/2^-$) are
illustrated in Fig.~\ref{fig2}, where we have plotted
$\theta$-trajectories for various values of the variational parameter
$w$. For relatively small values of $w$ the trajectory is a smooth
curve. As $w$ is increased this curve gains a turning point that
eventually becomes a cusp (left-hand part of Fig.~\ref{fig2}). Further
increase of $\theta$ leads to loop-like trajectories (right-hand part
of the figure).

\begin{figure}
\centerline{\includegraphics[width=8cm,clip]{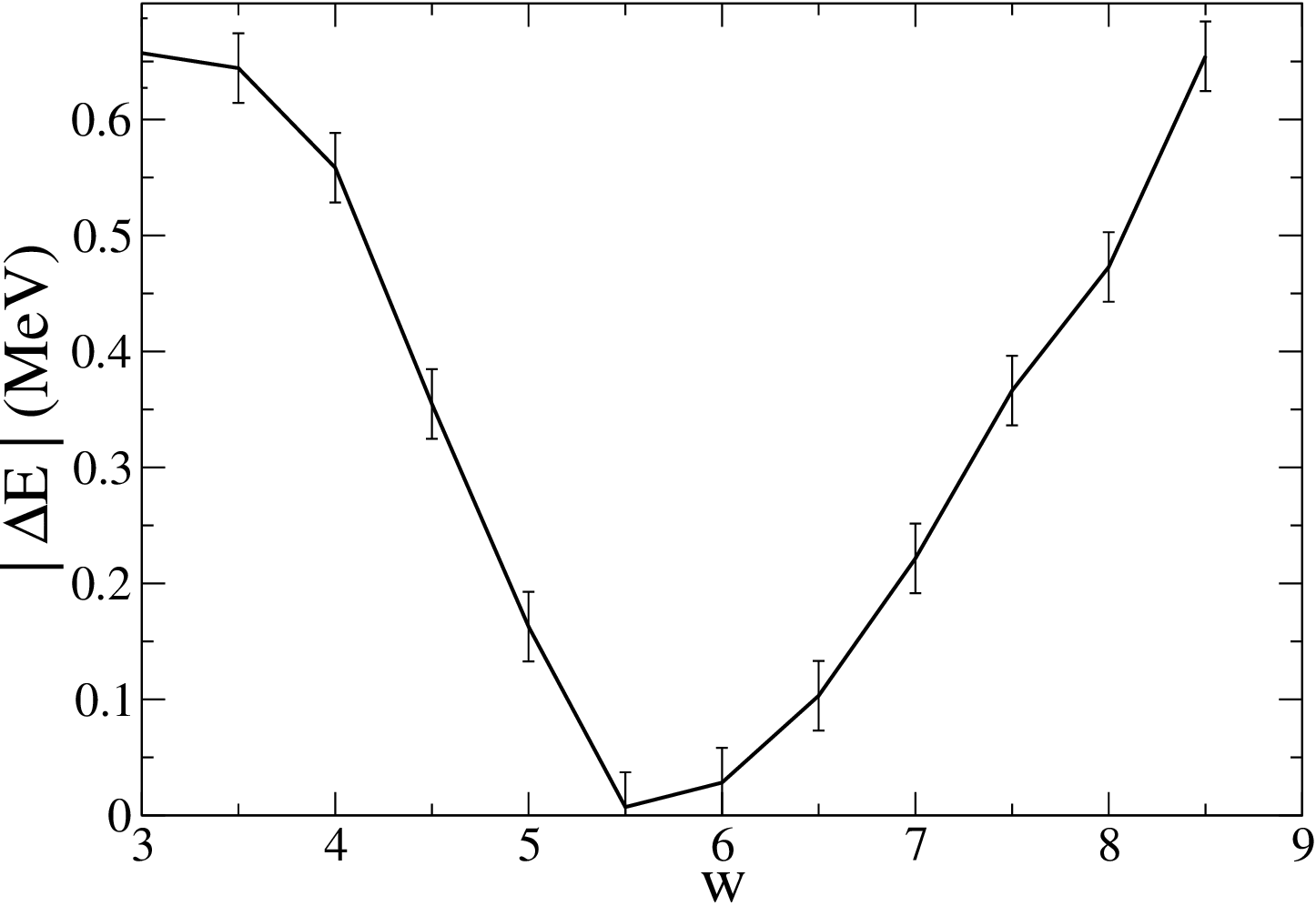}}
\caption{\label{fig3}The smallest absolute difference $|\Delta E|$
between adjacent values of the $\theta$-trajectory for a particular
value of $w$. The error bars are due to the Monte Carlo sampling. The
value of $w$ for which $|\Delta E|$ is minimum indicates the position
of the resonance.}
\end{figure}

In order to find the best $\theta$-trajectory we refer to the
derivative of the eigenvalue with respect to $\theta$. At resonance we
expect that the energy is stable with respect to $\theta$,
i.e. $|\frac{dE}{d\theta}|= 0$. Thus the best $w$ is that where
$|\frac{dE}{d\theta}|$ is closest to zero. This is illustrated in
Fig.~\ref{fig3}, where we can see the minimum value of $|\Delta E|$ of
the $\theta$-trajectory (corresponding to the absolute distance
between two adjacent points) for a particular value of $w$. The
presence of a resonance is observed as the minimum value of $|\Delta
E|$ that occurs in a range of $w$ between 5 and 6. According to
Fig.~\ref{fig2} this corresponds to a cusp in the trajectory. The
threshold for the problem (which is not shown in this figure) follows
the trajectory for the $^4$He ground state, as shown Fig.~\ref{fig1},
closely. This suggests little influence from the halo structure, on
this state, and little influence from this state on the continuous
spectrum of $^5$He.  From the obtained $\theta$-trajectories we can
infer a complex eigenvalue corresponding to a resonance in the region
($E_r=-26\pm 0.5$ MeV,$E_i=-23.2\pm 0.5$ MeV), for $\theta=0.35$. At
this point the real part of the threshold has moved, but it has not
developped an appreciable imaginary part. This movement means that the
threshold subtraction is somewhat uncertain. Nevertheless, it seems
most sensible to use the $\theta=0$ $^4$He binding energy as
threshold, $-27.3$ MeV. We thus find that a resonance 1.3$\pm$0.5 MeV
above threshold with a width of $2E_i=-46.4\pm 1$ MeV. The uncertainty
arising from the anomalous movement of the threshold with $\theta$ is
unknown, but as argued above probably small.

Having ensured the applicability of the complex-scaling method to
$^5$He we proceed to the case of $^6$He. In the case of $^6$He we
experimentally have a Borromean nucleus: while $^5$He is unbound, the
$^6$He ground state is stable. This state lies only 0.973 MeV below
the threshold for decay into an alpha-particle and two neutrons
($^4$He + $2n$) and thus $^6$He is very weakly bound. The first
resonance of $^6$He ($J^\pi=2^+$) lies 1.797 MeV above the ground
state and has a strong decay to the $^4$He + $2n$ channel. We find no
bound states in our calculation due to the lack of spin-orbit force in
the simplified nucleon-nucleon interaction used. However, we can
examine the structure of low-energy resonances in this light nucleus. In
our model $^6$He is described by two neutrons added to the alpha
particle wavefunction, where the additional coordinates are specified
relative to the alpha-particle center-of-mass. Similar to the case of
$^5$He the two-neutrons are placed within a spherical shell, each
parameterized in terms of a shifted Gaussian with an additional term
describing the interaction of the two neutrons. This leads to a
reference function, $\Phi_0$, that has the form
\begin{equation}
\Phi_0={\cal A}\{\Phi_\alpha{\cal Y}^L(r_{a5},r_{a6})
\exp\left(-\left(\frac{r_{a5}-d_5}{w_5}\right)^2\right)\exp\left(-\left(\frac{r_{a6}-d_6}{w_6}\right)^2\right)
\exp\left(-\left(\frac{r_{56}-d_{56}}{w_{56}}\right)^2\right)\},
\end{equation}
where $r_{ai}$ is the relative distance of the $i$th neutron from the
alpha-particle, while $r_{56}$ is the relative distance between the
two neutrons. ${\cal Y}^L(r_{a5},r_{a6})$ is a solid harmonic that
couples the two neutrons relative to the alpha particle to a total
orbital momentum $L$.

\begin{figure}
\centerline{\includegraphics[width=8cm,clip]{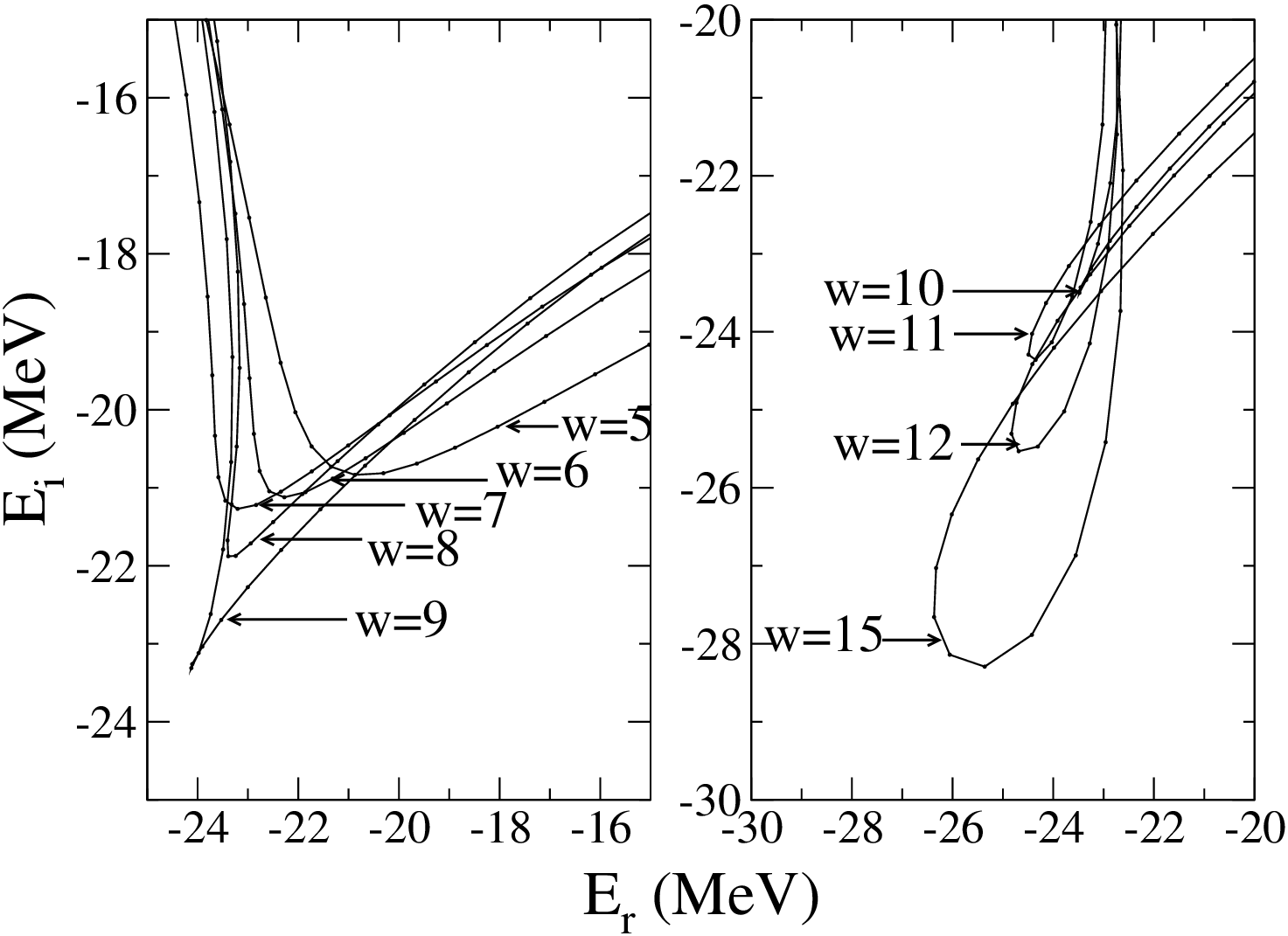}}
\caption{\label{fig4} The results obtained
for $^6$He $L=0$ ($J^\pi=0^+$) for various values of the width
parameter $w$. In this case $w$ is related to the separation of the
two neutrons that are placed in a shell around the alpha-particle. In
each plot the rotation parameter $\theta$ is varied in equal steps and
the $\theta$-trajectories evolve from right to left.}
\end{figure}

The calculation is considerably more difficult than that of $^5$He,
because of the increase both in configuration space and in the number
of variational parameters. However, it is sufficient to consider a
restricted set of variational parameters for a qualitative
understanding.  The results obtained look rather similar to those of
$^5$He. In Fig.~\ref{fig4}, the separation between the
alpha-particle and the dineutron center-of-mass is kept fixed, while
the the separation of the two-neutrons, represented by $w$, is
varied. For small values of $w$ the $\theta$-trajectories are smooth
curves, while a cusp is observed when $w$ is sufficiently
large. Again, this eventually changes into a loop.

Despite the qualitative nature of the results a clear resonant
structure is obtained. In both $^5$He and $^6$He calculations
only 9 components are used for the expansion of the correlation
operator, and this is enough for providing convergence. The small
basis set allows the calculations to be carried without any
significant restrictions in the numerical method. The only
difficulty that arises in going to heavier systems will be due to
the explicit antisymmetrization. As a result of the hermitian
representation of the complex variational principle is simple
to obtain  $\theta$-trajectories for the complex eigenvalues. In
order to identify the resonance we use the condition that at
resonance the complex energy should be stationary with respect to
the scaling parameter. For the moment the results obtained are somewhat 
qualitative. Although the numerical error due to the
Monte-Carlo sampling is trivial for a particular point in the
$\theta$-trajectories, the optimal value of the variational
parameters (e.g. $w$) that gives the resonance could only be
obtained within a range. Precise determination of that would
require a much more elaborate calculation.

Another point that needs further investigation is the fact that the
basis set employed does not provide a fixed threshold for the
alpha-particle. It is expected that this will further influence the
position of the resonance but is not clear at this stage in exactly
what way. Nevertheless, the difference between the
$\theta$-trajectories for the alpha-particle and those of $^5$He and
$^6$He is quite profound. The position where the resonance occurs for
both $^5$He and $^6$He is at a point where the motion of the
alpha-particle continuum due to the scaling is clockwise into the
lower complex plain. Therefore, it is extremely unlikely that the
resonance can be attributed to the anomalous behavior of the
alpha-particle threshold.


We have examined the application of a fully microscopic cluster model
to resonances in halo nuclei. We have obtained low-lying resonances
for both $^5$He and $^6$He by determining the point where the complex
energy is stationary with respect to a scaling parameter. This is
encouraging since experimentally it is expected that both systems have
low-lying resonances.

The central outcome of our investigation is that the
complex-scaling method can be successfully applied to our fully
microscopic model for weakly bound nuclei. This was achieved with
a relatively simple approximation, where the number of basis
functions is quite small. Within our framework of a non-orthogonal
expansion it is convenient to make use of the variational Eq.
(\ref{eq2}), rather than the standard complex variational
principle. Although both methods give the same results, the
hermitian representation provides a systematic way of obtaining
individual $\theta$-trajectories.

Apart from resonances our method provides an interesting way to
investigate the correlation mechanism used, which is widely applied
(Jastrow-CI). Despite the fact that such a correlation mechanism gives
a good description of the bound-state properties of the
alpha-particle, the ground state is not stationary under complex
scaling. This suggests that the basis used to expand the correlation
mechanism can be improved.

It is possible to obtain more accurate positions for these resonances
and in the future we plan to improve and extend the calculation in
order to get results of a more quantitative nature. This would include
both increasing the accuracy (more computer time), and obtaining upper
and lower bounds (as suggested in \cite{moiseyev:mo1,moiseyev:mo2}).

In this paper we are only interested in a qualitative understanding of
the method. However, a more complete interaction can be used, which
should lead to results comparable to the experimental data.  Hence,
the first natural step in extending the above work is to include a
spin-orbit force in the nucleon-nucleon interaction. It is expected
that this will be adequate to produce a bound state for $^6$He and at
the same time allow the results for resonances to be of a more
quantitative nature.

\section*{Acknowledgments}
This research was supported the EPSRC under grant GR/N15665. We wish to
thank A. Csoto for a discussion of his work.

\end{document}